\documentclass[proceedings,opts]{JHEP}
\conference{Heavy Flavours 8, Southampton, UK, 1999}
\usepackage{epsfig}

\def\etal{{\it et al.}}

\title{Future of Heavy Flavour Physics: Experimental Perspective}
\author{Sheldon Stone\\
{Physics Department, Syracuse University, Syracuse N. Y., 
USA, 13244-1130}\\
{E-mail: {\email{stone@phy.syr.edu}}}}

\abstract{I discuss what measurements need to be done to
search for physics beyond the Standard CKM model,
rather than just what studies can be done in the near future. It is also important to
accurately measure the CKM matrix elements.
Current best estimates for  two important elements are: 
 $|V_{cb}|=0.0381\pm 0.0021$ and $|V_{ub}/V_{cb}|=0.085\pm 0.019$. Finally,
future experiments are discussed.
 }

\begin{document}

\section{Introduction}

Our goals are to make an exhaustive search for physics beyond the Standard Model
and to precisely measure Standard Model parameters. 
Here we ask what studies need to be done, not just what studies can be done
in the near future. 
Measurements are necessary on CP violation in $B^o$ and $B_s$ mesons, $B_s$
mixing, rare $b$ decay rates, and mixing, CP violation and rare decays in the
charm sector.
These quarks were present in the early Universe. There is a connection between
our studies and Cosmology.

\section{The CKM Matrix and CP Violation}
\label{sec:Intro}
\subsection{The 6 Unitarity Triangles}
The base states of quarks, the mass eigenstates, are mixed to form the
weak eigenstates (primed) as described by the
Cabibbo-Kobayashi-Maskawa matrix, $V_{CKM}$ \cite{ckm},
\begin{equation}
\left(\begin{array}{c}d'\\s'\\b'\\\end{array} \right) =
\left(\begin{array}{ccc} 
V_{ud} &  V_{us} & V_{ub} \\
V_{cd} &  V_{cs} & V_{cb} \\
V_{td} &  V_{ts} & V_{tb}  \end{array}\right)
\left(\begin{array}{c}d\\s\\b\\\end{array}\right)~.
\end{equation}

There are 9 complex CKM elements. These 18 
numbers can be reduced to 4 independent quantities by applying unitarity 
constraints and the fact that the phases of the quark wave functions are 
arbitrary. 
These 4 remaining numbers are  fundamental constants of nature that 
need to be determined from experiment, like any other
fundamental constant such as $\alpha$ or $G$. In the Wolfenstein 
approximation \cite{wolf} $V_{CKM}$ equals:
\begin{equation}
{\scriptsize\begin{array}{ccc} 
1-\lambda^2/2 &  \lambda & A\lambda^3(\rho-i\eta(1-\lambda^2/2)) \\
-\lambda &  1-\lambda^2/2-i\eta A^2\lambda^4 & A\lambda^2(1+i\eta\lambda^2) \\
A\lambda^3(1-\rho-i\eta) &  -A\lambda^2& 1  
\end{array}}.
\end{equation}
This expression is accurate to order $\lambda^3$ in the real part and
$\lambda^5$ in the imaginary part. It is necessary to express the matrix
to this order to have a complete formulation of the physics we wish to pursue.
The constants $\lambda$ and $A$ have been measured as approximately 0.22 and
0.8, respectively, using semileptonic
$s$ and $b$ decays \cite{virgin}. Constraints on $\rho$ and $\eta$ exist from
other measurements.

Non-zero $\eta$ allows for CP violation. 
CP violation thus far has only been seen in the neutral kaon 
system. If we can find CP violation in the $B$ system we could see
if the CKM  model works or perhaps discover new physics that 
goes  beyond the model, if it does not.

\begin{figure}[htb]
\centerline{\epsfig{figure=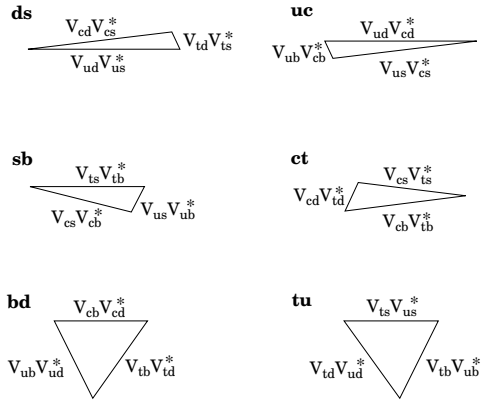,height=2.1in}}
\caption{\label{six_tri}The six CKM triangles. The bold labels, e.g. {\bf ds} 
refer to the rows or columns used in the unitarity relationship.}
\end{figure}

The unitarity of the CKM matrix allows us to construct six relationships.
These equations may be thought of triangles in the complex plane. They are 
shown in Figure~\ref{six_tri}.

All six of these triangles can be constructed knowing four and
only four independent angles \cite{silva_wolf} \cite{KAL} \cite{bigis}.
 These phases 
are taken as:
\begin{eqnarray} \label{eq:chi}
\beta=arg\left(-{V_{tb}V^*_{td}\over V_{cb}V^*_{cd}}\right),&~~&
\gamma=arg\left(-{{V^*_{ub}V_{ud}}\over {V^*_{cb}V_{cd}}}\right), \nonumber\\
\chi=arg\left(-{V^*_{cs}V_{cb}\over V^*_{ts}V_{tb}}\right),&~~&
\chi'=arg\left(-{{V^*_{ud}V_{us}}\over {V^*_{cd}V_{cs}}}\right).\nonumber\\ 
\end{eqnarray}

Two of the phases $\beta$ and $\gamma$ are probably large while $\chi$ is
estimated to be small $\approx$0.02, but measurable, while $\chi'$ is likely
to be one order of magnitude smaller than $\chi$.

\begin{figure}[htb]
\centerline{\epsfig{figure=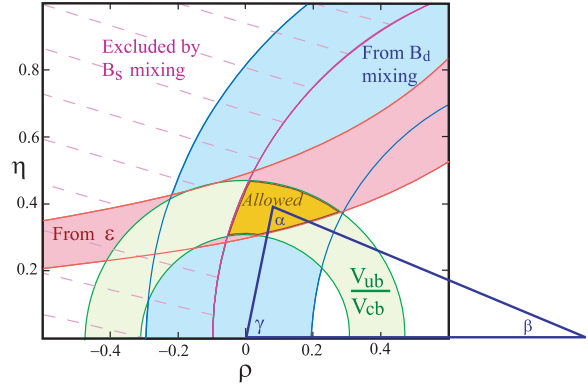,height=2in}}
\vspace{-0.2cm}
\caption{\label{ckm_tri}The CKM triangle shown in the $\rho-\eta$ plane. The
shaded regions show $\pm 1\sigma$ contours given by
$|V_{ub}/V_{cb}|$, neutral $B$ mixing, and CP violation in $K_L^o$ decay ($\epsilon$). The dashed region is excluded by $B_s$ mixing limits.
 The allowed region is defined by the overlap of
the 3 permitted areas, and is where the apex of the CKM triangle  sits.}
\end{figure}

In the {\bf bd} triangle, the one usually considered, the angles are all thought 
to be relatively large.
Since $V_{cd}^*=\lambda$, this triangle has sides
\begin{eqnarray}
1 & & \\ 
\left|{V_{td}\over A\lambda^3 }\right| &=& \sqrt{\left(\rho-
1\right)^2+\eta^2}
={1\over \lambda} \left|{V_{td}\over V_{ts}}\right|\\
\left|{V_{ub}\over A\lambda^3}\right|  &=& \sqrt{\rho^2+\eta^2}
={1\over \lambda} \left|{V_{ub}\over V_{cb}}\right|.
\end{eqnarray}
This CKM triangle is depicted in Figure~\ref{ckm_tri}, with
constraints from other measurements that will be discussed later. 

We know two sides already:
the base is  defined as unity and the left side is determined by 
measurements of  $|V_{ub}/V_{cb}|$. The right side can be determined using
mixing measurements in  the neutral $B$ system. There is, however,
 a large error due to  the  uncertainty in $f_B$, the $B$-meson
decay constant. This error can be greatly reduced by also measuring $B_s$
mixing.
The figure also shows the angles
$\alpha,~\beta$, and $\gamma$. Since they form a triangle the ``real" $\alpha$,
$\beta$ and $\gamma$ must sum to 180$^{\circ}$; therefore measuring any two
of these determines the third.

It has been pointed out by Silva and Wolfenstein \cite{silva_wolf} that
measuring these angles may not be sufficient to detect
new physics. For example, suppose there is new physics that arises in 
$B^o-\overline{B}^o$ mixing. Let us assign a phase $\theta$ to this new
physics. If we then measure CP violation in $B^o\to J/\psi K_S$ and eliminate
any Penguin pollution problems in using $B^o\to\pi^+\pi^-$, then we actually
measure $2\beta' =2\beta + \theta$ and $2\alpha' = 2\alpha -\theta$. So while
there is new physics, we miss it, because
$2\beta' + 2\alpha' = 2\alpha +2\beta$ and $\alpha' + \beta' +\gamma
= 180^{\circ}$.

\subsection{Ambiguities}
In measuring CP phases there are always ambiguities. For example, any 
determination of 
$\sin(2\phi)$, has a four-fold ambiguity; $\phi$, 
$\pi/2-\phi$, $\pi+\phi$, $3\pi/2-\phi$ are all allowed solutions. Often  
the point of view taken is that we know $\eta$ is a positive quantity and thus we 
can eliminate two of the four possibilities. However, this would be dangerous in 
that it could lead to our missing new physics. The only evidence that $\eta$ is 
positive arises from the measurements of $\epsilon$ and $\epsilon'$ and the fact 
that theoretical calculations give $B_K>0$ for $\epsilon$. Even accepting that
$K_L$ decays give $\eta >0$, it would be foolhardy to miss new physics just because 
we now assume that $\eta$ must be positive rather than insisting on a clean 
measurement of the angles that could show a contradiction.

\subsection{Technique for Measuring $\alpha$}
\begin{figure}[htb]
\vspace{-.3cm}
\centerline{\epsfig{figure=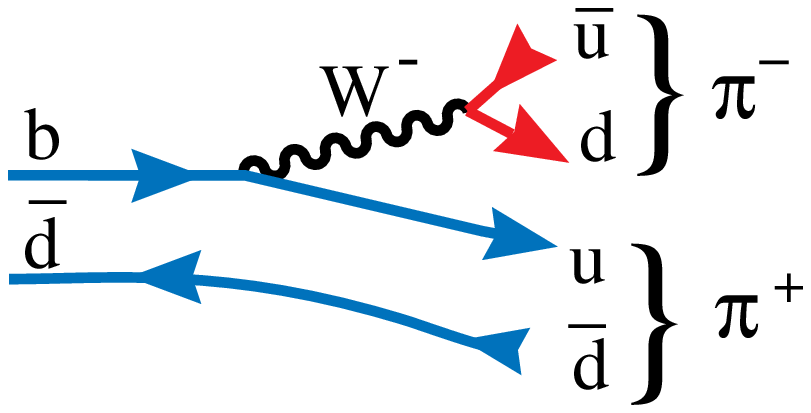,height=1.05in}\epsfig{figure=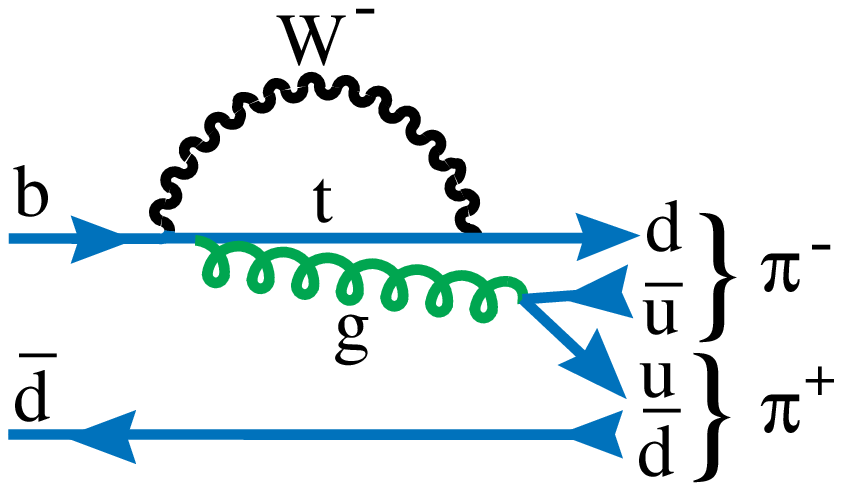,height=1.05in}}
\vspace{-0.2cm}
\caption{\label{pipi} Processes for $B^o\to\pi^+\pi^-$: Tree (left)
and Penguin (right).}
\end{figure}

It is well known that $\sin (2\beta)$ can be measured without
problems caused by Penguin processes using the reaction $B^o\to J/\psi K_S$.
The simplest reaction that can be used to measure $\sin (2\alpha)$ is
$B^o\to \pi^+\pi^-$. This reaction can proceed via both the Tree and Penguin
diagrams shown in Figure~\ref{pipi}.

 Current CLEO results are 
${\cal{B}}(B^o\to K^{\mp}\pi^{\pm})=(1.88^{+0.28}_{-0.26}\pm 0.13)\times 10^{-5}$ and 
 ${\cal{B}}(B^o\to \pi^{+}\pi^{-})=(0.47^{+0.18}_{-0.15}\pm 0.06) \times 10^{-5}$
\cite{wurthwein}, showing a relatively large Penguin amplitude that cannot
be ignored.  The
 Penguin contribution to $\pi^+\pi^-$ is roughly half the Tree
 {\it amplitude}. Thus the effect of the Penguin must be
determined in order to extract $\alpha$. The only model independent way 
of doing this was suggested by Gronau and London, but requires the measurement
of $B^{\mp}\to\pi^{\mp}\pi^o$ and $B^o\to\pi^o\pi^o$, the latter being rather 
daunting.

There is however, a theoretically clean method to determine $\alpha$.
The interference between Tree and Penguin diagrams can be exploited by
 measuring the time dependent CP violating
 effects in the decays $B^o\to\rho\pi\to\pi^+\pi^-\pi^o$  
as shown by Snyder and Quinn \cite{SQ}.

\begin{figure}[htb]
\vspace{-0.4cm}
\centerline{\epsfig{figure=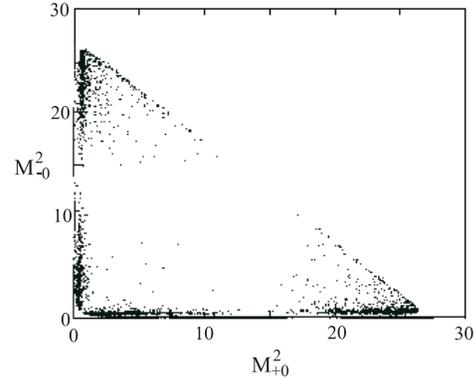,height=2.3in}}
\vspace{-.6cm}
\caption{\label{dalitz} The Dalitz plot for $B^o\to\rho\pi\to\pi^+\pi^-\pi^o$
from Snyder and Quinn.}
\end{figure}

The $\rho\pi$ final state has many advantages. First of all,
it has been seen with a relatively large rate. The 
branching ratio for the $\rho^o\pi^+$ final state as measured by CLEO is 
$(1.5\pm 0.5\pm 0.4)\times 10^{-5}$, and the rate for the neutral
 $B$ final state $\rho^{\pm}\pi^{\mp}$ is  
$(3.5^{+1.1}_{-1.0}\pm 0.5)\times 10^{-5}$, while the $\rho^o\pi^o$ final
state is limited at 90\% confidence level to $<5.1 \times 10^{-6}$
\cite{CLEO_rhopi}.  These
measurements are consistent with some theoretical expectations \cite{ali_rhopi}.
Furthermore, the associated vector-pseudoscalar
Penguin decay modes have conquerable or smaller branching ratios. Secondly, since the 
$\rho$ is spin-1, the $\pi$ spin-0 and the initial $B$ also spinless, the $\rho$ 
is fully polarized in the (1,0) configuration, so it decays as $cos^2\theta$, 
where $\theta$ is the angle of one of the $\rho$ decay products with the other
$\pi$ 
in the $\rho$ rest frame. This causes the periphery of the Dalitz plot to be 
heavily populated, especially the corners. A sample Dalitz plot is shown in 
Figure~\ref{dalitz}. This kind of distribution is good for maximizing the interferences, which 
helps minimize the error. Furthermore, little information is lost by excluding 
the Dalitz plot interior, a good way to reduce backgrounds.

To estimate the required number of events  Snyder and 
Quinn preformed an idealized analysis that showed that a background-free,
flavor-tagged sample of 1000 to
2000 events was sufficient. The 1000 event sample usually yields good results 
for $\alpha$, but sometimes does not resolve the ambiguity. With the 2000 event 
sample, however, they always succeeded. 

This technique not only finds $\sin(2\alpha)$, it also determines 
 $\cos(2\alpha)$, thereby removing two of the remaining ambiguities. The final
 ambiguity can be removed using the CP asymmetry in $B^o\to\pi^+\pi^-$ and
 a theoretical assumption \cite{gross_quinn}.

\subsection{Techniques for Measuring $\gamma$}

In fact, it may be easier to measure $\gamma$ than $\alpha$. There have been
at least four methods suggested.

(1) Time dependent flavor tagged analysis of $B_s\to D_s^{\pm}K^{\mp}$. This
is a direct model independent measurement \cite{Aleks}.

(2) Measure the rate differences between $B^-\to \overline{D}^o K^-$ and
$B^+\to {D}^o K^+$ in two different $D^o$ decay modes such as $K^-\pi^+$
and $K^+ K^-$. This method makes use of the interference between the tree
and doubly-Cabibbo suppressed decays of the $D^o$, and does not depend
on any theoretical modeling \cite{sad}\cite{gronau}.

(3) Rate measurements in two-body $B\to K \pi$ decays. A cottage industry has
developed. However, all methods are model dependent \cite{Kpi}.

(4) Use U-spin symmetry to relate $B^o\to\pi^+\pi^-$
and $B_s\to K^+ K^-$ \cite{pipiKK}.

\subsection{Required Measurements Involving $\beta$}

The phase of $B^o-\overline{B^o}$ mixing will soon be measured by
$e^+e^-$ $b$-factories using the $J/\psi K_S$ final state. New physics could be
revealed by measuring other final states such as $\phi K_S$, $\eta' K_S$
or $J/\psi \pi^o$. 

It is also important to resolve the ambiguities. There are two suggestions on how
this may be accomplished. Kayser \cite{kkayser} shows that time dependent
measurements of the final state
$J/\psi K^o$, where $K^o\to \pi \ell \nu$, give a direct measurement of
$\cos(2\beta)$ and can also be used for CPT tests. Another suggestion is to use
the final state $J/\psi K^{*o}$, $K^{*o}\to K_S\pi^o$, and to compare with
$B_s\to J/\psi\phi$ to extract the sign of the strong interaction phase shift
assuming SU(3) symmetry, and thus determine $\cos(2\beta)$ \cite{isi_beta}.

\subsection{A Critical Check Using $\chi$}

The angle $\chi$, defined in equation~\ref{eq:chi}, can be extracted by
measuring the time dependent CP violating asymmetry in the reaction
$B_s\to J/\psi \eta^{(}$$'^{)}$, or if one's detector is incapable of quality
photon detection the $J/\psi\phi$ final state can be used.  However, there are
two vector particles in the final state, making this a state of mixed CP
a requiring a complicated time-dependent angular analysis to find $\chi$.

Measurements of the magnitudes of 
CKM matrix elements all come with theoretical errors. Some of these are hard 
to estimate; we now try and view realistically how to combine CP violating phase 
measurements with the magnitude measurements to best test the Standard Model.

The best measured magnitude is that of $\lambda=|V_{us}/V_{ud}|=0.2205\pm 
0.0018$. Silva and 
Wolfenstein \cite{silva_wolf}, along with Aleksan, Kayser and London \cite{KAL}
show that the Standard Model 
can be checked in a profound manner by seeing if:
\begin{equation}
\sin\chi = \left|{V_{us}\over 
V_{ud}}\right|^2{{\sin\beta~\sin\gamma}\over{\sin(\beta+\gamma)}}~~.
\end{equation}
Here the precision of the check will be limited initially by the measurement of $\sin\chi$, not 
of $\lambda$. This check can  reveal new physics, even 
if other checks have not shown any anomalies. 

Other relationships to check include:
\begin{equation}
\sin\chi = \left|{V_{ub}\over 
V_{cb}}\right|^2{\sin\gamma~\sin(\beta+\gamma)\over{\sin\beta}}~~,
\end{equation}
\begin{equation}
\sin\chi = \left|{V_{td}\over 
V_{ts}}\right|^2{\sin\beta~\sin(\beta+\gamma)\over{\sin\gamma}}~~.
\end{equation}

The astute reader will have noticed that these two equations lead to the
non-trivial constraint:
\begin{equation}
\sin^2\beta\left|V_{td}\over V_{ts}\right|^2 = \sin^2\gamma\left|{V_{ub}\over 
V_{cb}}\right|^2 ~~.
\end{equation}
This contrains these two magnitudes in terms of two of the angles. 
Note, that it is in principle possible to determine the magnitudes of 
$|V_{ub}/V_{cb}|$ and $|V_{td}/V_{ts}|$ without model dependent errors
by measuring
$\beta$, $\gamma$ and $\chi$ accurately. Alternatively, $\beta$, $\gamma$
 and $\lambda$ can be used to give a much more precise value than is
possible at present with direct methods. For example, once $\beta$ and
$\gamma$ are known
\begin{equation}
\left|{V_{ub}\over V_{cb}}\right|^2 = \lambda^2
{{\sin^2\beta}\over {\sin^2(\beta +\gamma)}}~ ~~.
\end{equation}

\subsection{Other Critical CKM Measurements and Summary}

Magnitudes of the CKM elements are important to measure as precisely as possible.
Current measurements of $|V_{cb}|$ and $|V_{ub}|$ are discussed in 
section~\ref{sec:Vxy}.

It has been predicted that $\Delta\Gamma/\Gamma$ for the $B_s$ system is of
the order
of 10\%. This can be determined by measuring the lifetimes in different
final states such as $D_s^-\pi^+$ (mixed CP), $J/\psi \eta'$ (CP $-$) and
$K^+ K^-$ (CP +). A finite $\Delta\Gamma$ would allow many other interesting
measurements of CP violation \cite{DGamma}.

Table~\ref{table:reqmeas} lists the most important physics quantities and the
suggested decay modes. The necessary detector capabilities include
the ability to collect purely hadronic final states, the ability to identify
charged hadrons, the ability to detect photons with good efficiency and
resolution and  excellent time resolution required to analyze rapid $B_s$
oscillations.

\begin{table}[hbt]
\centering
\caption{Required Measurements for $b$'s}
\label{table:reqmeas}
\vspace*{2mm}
\begin{tabular}{|l|l|} \hline
Physics & Decay Mode  \\
Quantity&             
                   \\\hline
$\sin(2\alpha)$ & $B^o\to\rho\pi\to\pi^+\pi^-\pi^o$ \\
$\cos(2\alpha)$ & $B^o\to\rho\pi\to\pi^+\pi^-\pi^o$ \\
sign$(\sin(2\alpha))$ & $B^o\to\rho\pi$ \& $B^o\to\pi^+\pi^-$ \\
$\sin(\gamma)$ & $B_s\to D_s^{\pm}K^{\mp}$ \\
$\sin(\gamma)$ & $B^-\to \overline{D}^{0}K^{-}$  \\
$\sin(\gamma)$ & $B^o\to\pi^+\pi^-$ \& $B_s\to K^+K^-$  \\
$\sin(2\chi)$ & $B_s\to J/\psi\eta',$ $J/\psi\eta$\\
$\sin(2\beta)$ & $B^o\to J/\psi K_S$ \\
$\cos(2\beta)$ &  $B^o\to J/\psi K^o$, $K^o\to \pi\ell\nu$  \\
$\cos(2\beta)$ &  $B^o\to J/\psi K^{*o}$ \& $B_s\to J/\psi\phi$  \\
$x_s$  & $B_s\to D_s^+\pi^-$ \\
$\Delta\Gamma$ for $B_s$ & $B_s\to  J/\psi\eta'$, $ D_s^+\pi^-$, $K^+K^-$  \\
\hline
\end{tabular}
\end{table}

\section{Searches for New Physics}

Because new physics at much larger mass scales can appear in loops, rare
process such as $b\to s\gamma$, $d\gamma$, $s\ell^+\ell^-$ and
$d\ell^+\ell^-$ have the promise to reveal new physics. Searches in both
exclusive and inclusive final states are important.

Charm decays also offer the possibility of finding new physics in the study of
either mixing or CP violation as the Standard Model prediction is small.
The current experimental measurement of mixing is $r_D < 5\times 10^{-3}$,
while the SM expectation is $10^{-7}-10^{-6}$ \cite{charm_mix}. For CP violation the current
limits are about 10\%, while the expectation is $10^{-3}$ \cite{charm_cp}.

\section{Current Values of $|V_{cb}|$ and $|V_{ub}|$, and Allowed Regions
in $\rho-\eta$ Plane.}\label{sec:Vxy}
\subsection{Measurement Of $|V_{cb}|$ Using $B\to D^*\ell\nu$}
\begin{figure}
\vspace{-17mm}
\centerline{\epsfig{figure=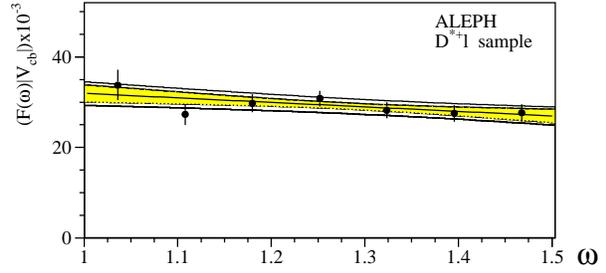,width=3.8in}}
\vspace{-42mm}
\caption{\label{aleph_vcb} $\overline{B}^o\to D^{+}\ell^-\bar{\nu}$ from ALEPH. 
The data have been
fit to a functional form suggested by Caprini \etal ~The abcissa gives the
value of the product $|F(\omega)*V_{cb}|$.}
\end{figure}

Currently, the most favored
technique is to measure the decay rate of $B\to D^{*}\ell^-\bar{\nu}$ at the
kinematic point where the $D^{*+}$ is at rest in the $B$ rest frame (this is 
often referred to as maximum $q^2$ or $\omega =1$). Here, according to Heavy
Quark Effective Theory, the theoretical
uncertainties are at a minimum. The ALEPH results \cite{aleph_pi} are shown in
Figure~\ref{aleph_vcb}.

Table~\ref{tab:Vcb} summaries determinations of $|V_{cb}|$; here,
the first error is statistical, the second systematic and the third, an estimate 
of the theoretical accuracy in predicting the form-factor $F(\omega=1)=0.91\pm
0.03$ \cite{formfactor}. The value and accuracy have been questioned 
\cite{bigivcb}.
Hopefully, in the near future a reliable value will be given by 
lattice QCD without using the quenched approximation \cite{simone}.
Currently, DELPHI has the smallest error, they detect  only the slow $\pi^+$
from the $D^{*+}$ decay and do not 
reconstruct the $D^o$ decay.
CLEO, however, has only used 1/6 of their current data. The quoted average $|V_{cb}|=0.0381\pm 0.0021$
combines the averaged statistical and systematic errors with the theoretical 
error in quadrature and
takes into account the common systematic errors, such as the $D^*$ branching
ratios.

\begin{table}[th]
\vspace{-2mm}
\begin{center}
\caption{Modern Determinations of $|V_{cb}|$ using 
$B\to D^*\ell^-\overline{\nu}$ decays
at $\omega = 1$ \label{tab:Vcb}}
\vskip 0.1 in
\begin{tabular}{|l|c|}\hline
Experiment & $V_{cb}$ $(\times 10^{-3})$\\
\hline
\hline
ALEPH\cite{aleph_pi} & $34.4\pm 1.6 \pm 2.3 \pm 1.4$ \\
DELPHI\cite{delphi_pi} & $41.2\pm 1.5 \pm 1.8 \pm 1.4$ \\
OPAL\cite{opal_pi} & $36.0\pm 2.1 \pm 2.1 \pm 1.2$ \\
CLEO\cite{cleo_pi} & $39.4\pm 2.1 \pm 2.0 \pm 1.4$ \\
\hline
Average & $38.1\pm 2.1$\\
 \hline
\end{tabular}
\end{center}
\end{table}
There are other ways of determining $V_{cb}$. One method based on QCD sum rules uses the 
operator product expansion and the heavy quark expansion, in terms of  the parameters $\alpha_s (m_b)$, $\overline{\Lambda}$, and the matrix elements $\lambda_1$ and $\lambda_2$. 
The latter quantities arise from the differences
\begin{displaymath}
m_B-m_b=\overline{\Lambda}-{{\lambda_1+3\lambda_2}\over{2m_b}}~~~
m_B^*-m_b=\overline{\Lambda}-{{\lambda_1-\lambda_2}\over{2m_b}}~~.
\end{displaymath}
 The $B^*-B$ mass difference determines $\lambda_2 = 0.12$ GeV$^2$.
The total semileptonic width is then related 
to these parameters \cite{parms}.

CLEO has measured the semileptonic branching ratio using lepton tags as 
(10.49$\pm$0.17$\pm$0.43)\% and using the world average lifetime for an equal 
mixture of $B^o$ and $B^-$ mesons of  1.613$\pm$0.020 ps, CLEO finds
$\Gamma_{sl} = 65.0\pm 3.0$ ns$^{-1}$. 

\begin{figure}[hbt]
\vspace{-3mm}
\centerline{\epsfig{figure=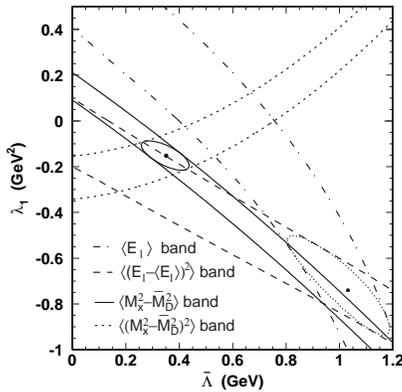,width=2.1in}}
\vspace{5mm}
\caption{\label{moments}Bands in $\overline{\Lambda}-\lambda_1$ space found
by CLEO in analyzing first and second moments of hadronic mass squared and
lepton energy. The intersections of the two moments for each set determines
the two parameters. The 1$\sigma$  error ellipses are shown.}
\end{figure}

CLEO then attempts to measure the remaining unknown parameters $\lambda_1$ 
and $\overline{\Lambda}$ by using moments of the either the hadronic mass or 
the lepton energy \cite{moment}. The results are shown in Figure~\ref{moments}. Here the 
measurements are
shown as bands reflecting the experimental errors. Unfortunately, this
preliminary CLEO result shows a contradiction. The overlap of the mass moment
bands gives different values than the lepton energy moments! The theoretically
favored mass
moments give the values $\lambda_1$=
(0.13$\pm$0.01$\pm$0.06) GeV$^2$, and  $\overline{\Lambda}$ = 
(0.33$\pm$0.02$\pm$0.08) GeV. The discrepancy between the two methods is
serious. It either means that there is something wrong with the CLEO analysis
 or there is something wrong in the theory, perhaps the breakdown of duality.
 If the latter is true it would shed
doubt on the method used by the LEP experiments to extract $|V_{ub}|$
using the same theoretical framework.

\subsection{Measurement Of $|V_{ub}|$}
Another important CKM element that can be measured using semileptonic decays is
$V_{ub}$. This is a heavy to light quark transition where HQET cannot be used.
Unfortunately the theoretical models that can be used to 
extract a value from the data do not currently give precise predictions. 

Three techniques have been used. The first measurement of $V_{ub}$ done by CLEO
 and subsequently
confirmed by ARGUS, used only leptons which were more energetic than those that
could come from $b\to c\ell^- \bar{\nu}$ decays \cite{first_vub}. These  
``endpoint leptons'' can occur $b\to c$ background free at the
$\Upsilon (4S)$, because the $B$'s are almost at rest. Unfortunately, there is
only  a small fraction of the $b\to u \ell^-\bar{\nu}$ lepton spectrum that
can be seen this way, leading to model dependent errors. The models used are
either inclusive predictions, sums of exclusive channels, or both
\cite{vubmods}. The average among the models is $|V_{ub}/V_{cb}|=0.079\pm 0.006$,
without a model dependent error.
These models differ by at most 11\%, making it tempting to assign a $\pm$6\%
error. However, there is no quantitative way of estimating the error.

\begin{figure}[bt]
\vspace{-9mm}
\centerline{\epsfig{figure=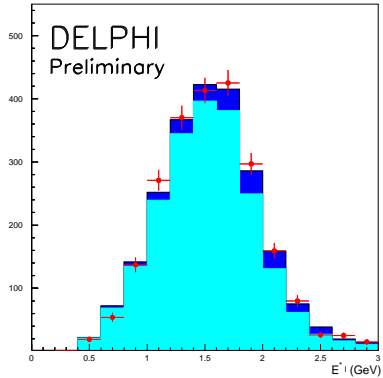,width=2.in}}
\vspace{-4mm}
\caption{\label{delphi_vub}The lepton energy distribution in the $B$ rest
frame from DELPHI. The data have been enriched in $b \to u$ events, and the 
mass of the recoiling hadronic system is required to be below 1.6 GeV. The 
points indicate data, the light shaded region, the fitted background and the 
dark shaded region, the fitted $b \to u \ell \nu$ signal. }
\end{figure}

ALEPH \cite{aleph_vub}, L3 \cite{L3_vub}
and DELPHI \cite{delphi_vub} try to isolate a class of events where the hadron 
system associated
with the lepton is enriched in $b\to u$ and thus depleted in $b\to c$.     
They define a likelihood that hadron tracks come from $b$ decay by using a large 
number of variables including, vertex information, transverse momentum, not 
being a kaon etc.. Then they require the hadronic mass to be less than 1.6 GeV, which 
greatly reduces $b\to c$, since a completely reconstructed $b\to c$ decay has a 
mass greater than that of the $D$ (1.83 GeV). They then examine the lepton 
energy distribution, shown in Figure~\ref{delphi_vub} for
DELPHI.

The average of all three results as given by the LEP working group
\cite{Lep_Vub} results
in $|V_{ub}/V_{cb}|=0.106^{+0.017}_{-0.020}$. The results use models
\cite{vub_thy_inc} \cite{bigivcb} that assume duality to extract the result. (I have used
$|V_{cb}|=0.0381\pm 0.0021$.) I have two grave misgivings about this result.
First of all the experiments have to understand the level of $b\to c\ell\nu$
backround to 0.6\%. They have not demonstrated that they can do this; there
are no experimental checks at this level. Secondly, the theory assumes duality,
and there are no successful experimental checks here either. The one possible
check, that of the $b\to c\ell\nu$ moments has not as yet succeeded. Therefore,
I choose not to use these results in my average.  
 

The third method uses exclusive decays.
CLEO has measured the decay 
rates for the exclusive final states $\pi\ell\nu$ and 
$\rho\ell\nu$ \cite{cleo_pirho}. The model of Korner and 
Schuler (KS) was ruled out by the measured ratio of $\rho/\pi$ \cite{vubmods}. 
CLEO has recently
presented an updated analysis for $\rho\ell\nu$ where
they have used several different models to evaluate
their efficiencies and extract $V_{ub}$. These
theoretical approaches include quark models, light cone sum
rules (LCRS), and lattice QCD. The CLEO values are shown
in Table~\ref{tab:Vub}.
\begin{table}[th]
\vspace{-2mm}
\begin{center}
\caption{Values of $|V_{ub}|$ using 
$B\to \rho\ell^-\overline{\nu}$ and some theoretical models \label{tab:Vub}}
\vskip 0.1 in
\begin{tabular}{|l|c|}\hline
Model & $V_{ub}$ $(\times 10^{-3})$\\
\hline
ISGW2\cite{vubmods} & $3.23\pm 0.14^{+0.22}_{-0.29}$ \\
Beyer/Melnikov\cite{BM} &$3.32\pm 0.15^{+0.21}_{-0.30}$  \\
Wise/Legeti\cite{WL} &$2.92\pm 0.13^{+0.19}_{-0.26}$  \\
LCSR\cite{LCSR} &$3.45\pm 0.15^{+0.22}_{-0.31}$  \\
UKQCD\cite{LCSR} &$3.32\pm 0.14^{+0.21}_{-0.30}$  \\
\hline
\end{tabular}
\end{center}
\end{table}

The uncertainties in the quark model calculations (first three in the table)
 are guessed to be
 25-50\% in the rate. The Wise/Ligetti model uses charm data and
SU(3) symmetry to reduce the model dependent errors. The other models estimate their errors
at about 30\% in the rate, leading to a 15\% error in $|V_{ub}|$. 
Note that the models differ by 18\%, but it would be incorrect to assume
that this spread allows us to take a smaller error. At this time it is
prudent to assign a 15\% model dependent error realizing that the errors
in the models cannot be averaged. The fact that the models do not differ
much allows us to comfortably assign a central value 
$|V_{ub}|=(3.25\pm 0.14^{+0.22}_{-0.29}\pm 0.50)\times 10^{-3}$, and
a derived value $|V_{ub}/V_{cb}|=0.085^{+0.008}_{-0.010}\pm 0.016$~. CLEO
could lower this error somewhat if the $\pi\ell\nu$ final state was reanalyzed
with LCSR and lattice gauge models. 
 
Only the lattice model predictions of UKQCD are used here. More
lattice gauge predictions for the rates in these reactions, at least in some
regions of $q^2$, are promised soon \cite{newlg} \cite{simone} with better errors.
My view is that with experimental checks from measuring form-factors 
and unquenched lattice gauge models the errors will eventually decrease.
 


We can use this estimate of $|V_{ub}/V_{cb}|$ along with other
measurements, to get some idea of what the likely values of $\rho$ and 
$\eta$ are.
The $\pm 1\sigma$ contours shown in Figure~\ref{ckm_tri}
 come from measurements of CP violation in $K_L^o$
decay ($\epsilon$), $|V_{ub}/V_{cb}|$ and $B^o$ mixing. Theoretical errors
dominate.  The limit on $B_s$ mixing also restricts the range;
its measurement is quite important. 
Some groups have tried to narrow the ``allowed region" by doing maximum
liklihood fits, assigning Gaussian errors to the estimated theoretical
parameters \cite{optimists}. I strongly disagree with this approach. The technique
of Plaszczynski, shown at this conference \cite{Plas}, while imprecise, is more justifiable.

\section{Future Experiments}
Lack of space precludes a more through review here. The $e^+e^-$ experiments,
BaBar, Belle and CLEO should see CP violation in 2000. The first two in
the $J/\psi K_S$ final state, while CLEO has a chance of seeing direct CP
violation in rare decays. CDF and D0 are now scheduled to  turn on in 2001.
CDF already has
seen some evidence for CP violating effects in $J/\psi K_S$ \cite{borto}, and
promises to measure $B_s$ mixing. HERA-B should also turn on in this period.

To overconstrain the CKM matrix and look for new physics all the quantities
listed in Table~\ref{table:reqmeas} requires, however, much larger samples
of $b$-flavored hadrons, and detectors capable of tolerating large interaction
rates and having excellent lifetime resolution, particle identification
and $\gamma/\pi^o$ detection capabilities. The large $b$ rates, including the
$B_s$, are available only at hadron colliders. Two dedicated experiments are
contemplated, LHC-b which has been approved and BTeV which is being proposed.
Harnew \cite{Harnew} has shown the prospects for these two experiments and ATLAS and CMS
in these proceedings. I only add that the PbWO$_4$ EM calorimeter of BTeV 
should provide important capabilities beyond LHC-b. 

\acknowledgments
I thank John Flynn and Chris Sachrajda for organizing an excellent and
fruitful meeting. I thank my colleagues Marina Artuso and Tomasz Skwarnicki
for many useful discussions.

\end{document}